# Performance Analysis of a Photovoltaic System with Thermoelectric Generator and Phase Change Material; An Experimental Approach

Tobechukwu Okamkpa[1, a], Joshua Okechukwu[1, b *], Divine Mbachu[1, 2, c] and Chigbo Mgbemene[1, d]

[1] *AEDJAC Lab, University of Nigeria, Nsukka 410001, Enugu, Nigeria*

[2] *Department of Electrical Engineering, Stanford University, Stanford, CA 94305, USA*

[a]tobechukwu.okamkpa.242531@unn.edu.ng, [b]joshua.okechukwu.247915@unn.edu.ng *, [c]mbachu@stanford.edu, [d]chigbo.mgbemene@unn.edu.ng

**Abstract.** This study explores the integration of thermoelectric generators (TEGs) and phase change materials (PCMs) to enhance the efficiency of photovoltaic (PV) panels in high-temperature conditions. An AP-PM-20 Polycrystalline PV panel, SP-1848-27145 Bismuth Telluride TEG, and paraffin wax PCM in an aluminum container were used. Four configurations were tested: standalone PV, PV-PCM, PV-TEG-PCM, and PV-PCM-TEG, under identical conditions from 10:30 AM to 6:00 PM at 25-minute intervals. Data on PV and TEG voltage, current, and solar irradiance were collected and analyzed. The results show significant performance improvements: the PV-PCM configuration boosted power output by 68.04%, while PV-PCM-TEG and PV-TEG-PCM configurations improved efficiency by 43.06% and 37.51%, respectively. Efficiency gains relative to the standalone PV system were 33.33% for PV-PCM, 25.76% for PV-PCM-TEG, and 21.21% for PV-TEG-PCM, demonstrating the effectiveness of PCMs and TEGs in enhancing PV performance.

## Introduction

**High-temperature Challenges in Photovoltaic Systems.** The present energy situation around the world emphasizes more on renewable sources for electricity production. Fossil fuels are increasingly seen as unsustainable for the future because of their non-renewable nature and significant contribution to environmental pollution. Out of all the available renewable options, photovoltaic (PV) energy has gained the most popularity [1]. This is mainly because the earth receives a vast amount of solar energy. Among all other renewable sources, PV technology has the highest power density, is more cost-effective, produces no pollutants during use, and requires minimal maintenance [2, 3]. All these attributes make solar PV electricity desirable. However, PV technology has its challenges despite its numerous benefits, especially with regard to surface operating temperature [4]. An increase in the solar radiation intensity and surrounding ambient temperature, mostly leads to a rise in the photovoltaic cells operating temperature. This results in a reduction in the overall power output of both polycrystalline and monocrystalline photovoltaic cells, which are widely used in many power applications [5]. It is therefore essential that PV systems become more efficient to ensure the viability of this renewable technology. Consequently, alternative approaches must be found for mitigating problems related to temperature in PV cells, thereby leading to the enhancement of conversion efficiency.

**Thermoelectric Cooling Solutions for Photovoltaic Systems.** Various studies have been carried out to find novel ways of cooling solar panels, and it was observed that an improvement in cooling strategies of solar systems can increase it conversion efficiency. A notable cooling system that has gained prominence is thermoelectric cooling systems that employ thermoelectric generators (TEGs) consisting of n- and p-type semiconductors arrayed thermally in parallel but electrically in series. TEGs utilize the temperature difference, caused by the Peltier effect, to produce increased electrical power while at the same time dissipating excess heat from PV panels [6]. Because they are simple, small, have no movable parts, and need little maintenance, thermoelectric devices are very attractive among waste heat recovery technologies [7]. There is also another innovative method that involves using phase-change materials (PCMs) to remove heat from PV panels. PCMs have latent heat storage characteristics that have proved effective in keeping PV panels cool [8, 9, 10].

Several research works have studied how thermoelectric generators (TEGs) can be used as heat sinks for PV panels. A model presented in [11] optimized TEG device shapes for PV/TE systems, demonstrating enhanced power output and efficiency through simulations. Similarly, [12] compared standalone solar and hybrid PV/TE systems, finding a 1.47% efficiency increase and 61.01% power boost for the PV-TE-heat pipe system at a 6:1 ratio over PV-only systems. A thermodynamic model for a concentrated PV-TE hybrid system in [13] showed a 14% power increase with TEG modules at an optimal concentration ratio of 5.5 kW/m². Studies in [14] optimized load resistance and PV output voltage to identify the optimal configuration for maximizing both temperature management and power output in a hybrid PV-TE system. In [15], a perovskite solar cell (PSC) and TEG hybrid system with enhanced thermal stability and photoelectric conversion, was introduced, achieving a 22% increase in photoelectric conversion efficiency compared to standalone PSCs. The system optimized at 1.87 V VOC under AM 1.5 G light and a 15°C temperature gradient. Similarly, [16] proposed a PV-TE system integrating polycrystalline cells, water cooling, and TEG modules. A 24-hour and one-year FEA simulation showed a 1.67% increase in electrical efficiency and 1.72% in exergy efficiency for the hybrid system compared to the standalone PV. In [17], the economic, thermodynamic, and environmental analysis of photovoltaic/thermal systems was conducted. Heat transfer analysis was done for various fluids such as: water, Al2O3 homogenous composition nanofluids, and combined Al2O3/Cu nanofluids of varying nanoparticle morphologies. While research by [18] suggests that photovoltaic/thermal (PV/T) systems with thermoelectric generators (TEGs) may outperform standalone PV/T systems in terms of energy transfer, efficiency, and environmental/economic feasibility, particularly with nanofluids like Al2O3/Cu, their effectiveness diminishes in hot desert regions with concentrated solar irradiation. In such environments, the performance degradation of the PV cell due to elevated temperatures outweighs the additional power generation from the TEG.

**Thermal Management Improvement in Photovoltaic Cells using Phase Change Materials.** PCMs, with their high latent heat capacity, effectively regulate PV panel temperatures through thermal storage [19]. Their application in thermal energy storage systems is expanding due to this property [20], making them suitable for PV-TEG systems. PCMs store excess solar energy for TEG use during low-radiation periods, enhancing system efficiency [21, 22, 23]. Studies in [24] compared active (TEG-only) and hybrid (TEG-PCM) cooling for PV panels. The hybrid system improved efficiency by 2.5-3.5% and increased power generation by 20-30% under different conditions. [25] explored integrating TEGs and PCMs in building-integrated photovoltaic (BIPV) systems to enhance power output. [26] proposed a novel heat sink with PCM and nanofluid, boosting PV/TEG system electrical efficiency by 12.28% and exergy efficiency by 11.6% compared to water-only cooling. Studies like [27] explored using copper foam and expanded graphite to improve existing PCM designs for concentrated PV-TEG systems and found that lowering the thermal resistance of PCMs increased TEG output without impacting photovoltaic operation. Similarly, [28] investigated synthesizing coconut oil and beeswax composites as PCMs to regulate PV panel temperature. Their findings showed that the PV/TEG-PCM system improved power output and efficiency while reducing temperature and stabilizing monocrystalline cells. Building on this, [29] explored a novel PV system with two PCMs and strategically placed metals for optimized heat transfer. This configuration outperformed a baseline PV/TEG system with a single PCM. An approach to a multi-dimensional, non-stationary - time dependent thermal simulation study of the PV integrated TEG-PCM hybrid system accounting for changes in solar radiation is described in [15]. Analysis of the data revealed that neither the standalone photovoltaic (PV) configurations nor the PV-TEG combinations exhibited better performance when compared to the PV cell integrated with TEG-PCM. However, this three-dimensional numerical model has yet to be experimentally validated under real weather conditions.

The aim of this paper is to verify and validate the improvements observed by numerical models in prior literature through experimentation of different configurations conducted under identical operating conditions.

**Research Methodology**

**Physical Model Description and Specifications.** The system's three-dimensional representation was created using Autodesk Fusion 360. Fig. 1a shows the standalone PV system with a solar panel that includes various components such as glass cover, ethylene vinyl acetate (EVA), silicon PV cell, and tedlar. This solar panel utilizes energy from the sun to produce electricity as specified in Table 1 and Table 4. Conversely, Fig. 1b illustrates an encapsulated phase change material placed directly below the PV panel. A 1[mm] thick aluminum sheet was reshaped to form a box measuring 45 ×50 ×10 [mm], which served as the PCM container. Paraffin wax is used as the phase change material to store excess heat energy from the solar panel to mitigate the effect of high cell temperature on the output of the PV cells. A 1 [mm] thick layer of thermal paste is applied between the PCM and the back surface of the PV panel to ensure effective thermal dispersion. The properties of the PCM are shown  shown in Table 2.

Table 1: PV panel specifications

| S/N | Specifications | |
|---|---|---|
| 1 | Type | Polycrystalline |
| 2 | Maximum Power ($P_{max}$) | 0.12 [W] |
| 3 | Current at ($I_{mp}$) | 60 [mA] |
| 4 | Voltage at ($V_{mp}$) | 2 [V] |

Figure 1c. shows the TEG placed below the PV panel to potentially utilize excess heat. The PV panel directs thermal energy to the hot side of the TEG, while the PCM on the cold side absorbs heat, creating a temperature gradient that allows for additional power generation via the thermoelectric effect. The TEG specifications are provided in Table 3.

Continuing the visual representation, as illustrated in Fig. 1d, the PV panel is stationary, and directly beneath it is the contained phase change material to store extra heat from the PV panel. The PCM gives out the stored heat to the thermoelectric generator so that it can be used for secondary purposes of energy generation. In order to help with this, a thermally conductive paste is normally used when attaching them at the lower side of PCM.  To further augment the temperature differential across the TEG, a heat sink is positioned adjacent to the cold side. This heat sink facilitates efficient heat dissipation to the surrounding environment, thereby maintaining a lower cold side temperature.

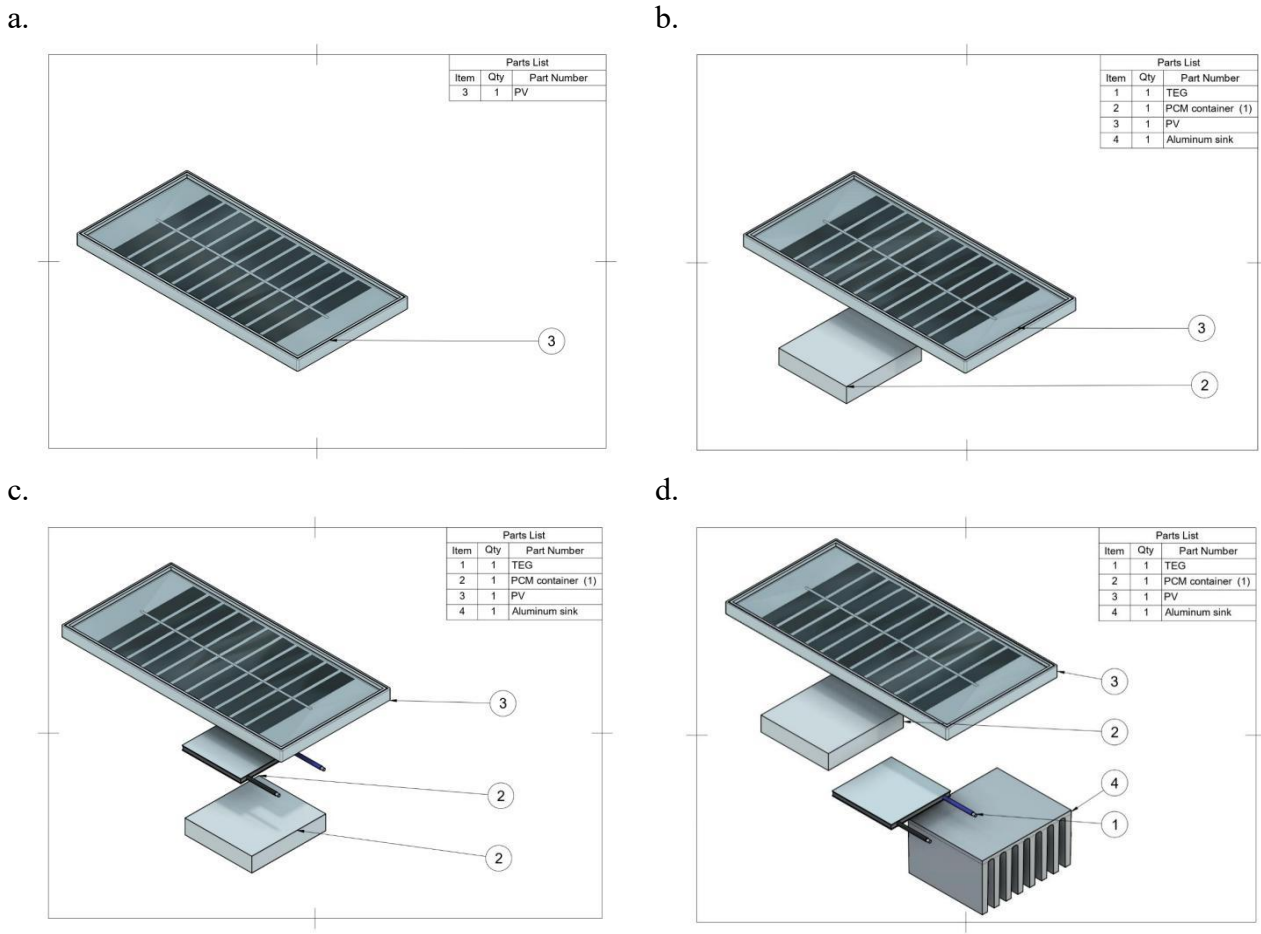

Fig 1. Configurations: (a) Stand-alone PV; (b) PV-PCM; (c) PV-TEG-PCM; (d) PV-PCM-TEG

Table 2: PCM specifications

| S/N | Specifications | |
|---|---|---|
| 1 | Material | Paraffin wax |
| 2 | Density at 25 [°C] | 810 [kg/m$^3$] |
| 3 | Volume | 0.0225 [litres] |
| 4 | Weight | 0.018225 [kg] |
| 5 | Volumetric thermal expansion coefficient | 0.0007 [kJkg$^{-1}$K$^{-1}$] |
| 6 | Fla[mm]ability temperature | > 300 [°C] |
| 7 | Thermal conductivity at [25°C] | 0.2 [[Wm-1K-1] |
| 8 | Boiling temperature | 334 [°C] |
| 9 | Latent heat of fusion | 210 [kJkg$^{-1}$] |
| 10 | Melting temperature | 45 [°C] |

Table 3: TEG specifications

| S/N | Specifications | |
|---|---|---|
| 1 | Material | $Bi_2Te_3$ |
| 2 | Dimensions | 40 x 40 x 3 [mm] |
| 3 | Weight | 25 [g] |
| 4 | Thermal conductivity | 0.797 [$Wm^{-1}K^{-1}$] |
| 5 | Voltage at 20 [°C] temperature difference | 0.97 [V] |
| 6 | Current at 20 [°C] temperature difference | 225 [mA] |
| 7 | Voltage at 40 [°C] temperature difference | 1.8 [V] |
| 8 | Current at [40°C] temperature difference | 368 [mA] |
| 9 | Voltage at [60°C] temperature difference | 2.4 [V] |
| 10 | Current at [60°C] temperature difference | 469 [mA] |
| 11 | Voltage at [80°C] temperature difference | 3.6 [V] |
| 12 | Current at [80°C] temperature difference | 559 [mA] |
| 13 | Voltage at [100°C] temperature difference | 4.8 [V] |
| 14 | Current at [100°C] temperature difference | 669 [mA] |

Table 4: Specifications of materials used

| S/N | Component | Dimensions | Necessity | Material | Model |
|---|---|---|---|---|---|
| 1 | PV | 120 [mm] × 75 [mm] × 10 [mm] | Conversion of solar energy to DC power | Polycrystalline | AP-PM-20 |
| 2 | TEG | 40 [mm] × 40 [mm] × 3 [mm] | Heat energy to DC conversion | Bismuth telluride $Bi_2Te_3$ | SP-1848-27145 |
| 3 | PCM | | Heat extraction on the cold side of the TEG | Paraffin wax | |
| 4 | Metallic container | 45 [mm] × 50 [mm] × 10 [mm] | Contains the phase change material | Aluminium | |

**Experimental Work Procedures.** All experiments were carried out at AEDJAC Lab, located on the rooftop of the Faculty of Engineering building, University of Nigeria, Nsukka, Enugu state, Nigeria (6.8664°N, 7.4093°E), November 2023. The lab has an unobstructed view of the sky suitable for solar energy research. In Fig. 2, the experimental setups for the different configurations are shown.

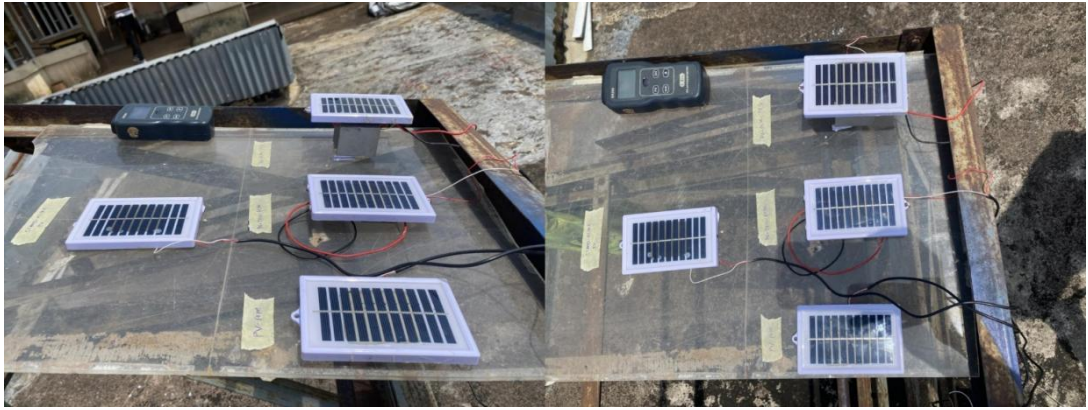

Figure 2: Experimental setup of the different configurations: Standalone PV, PV-PCM, PV-PCM-TEG, and PV-TEG-PCM systems.

During experimentation, the four setups: standalone PV, PV-PCM, PV-PCM-TEG, and PV-TEG-PCM, were all subjected to the same weather conditions. These setups were tested simultaneously using a solarimeter and multimeter attached to the different configurations to record data. Immediately after the setups were made, the data was collected for over 8 hours (10:30 am to 6:00 pm), with 25-minute intervals. This data collected includes PV voltage and current, TEG voltage and current, and solar irradiance value. The solar irradiance value was measured using a solarimeter, while the TEG and PV voltage and current were measured using multimeters of the same ratings.

**Mathematical Model of the PV System.** The solar panel converts the incoming sunlight into direct current (DC) electricity, and the temperature difference within the thermoelectric generator generates DC electricity. The individual efficiencies of these components and their collective efficiency can be determined using mathematical expressions.

**Photovoltaic panel.** For the PV panel, the efficiency ($\eta_{pv}$) after hours of exposure to sunlight can be calculated using Eq. (1) and Eq. (2)

$$\eta_{pv} = \frac{P_{pv}}{P_{in}} = \frac{I_{pv,max} \times V_{pv,max}}{P_{in}} \qquad (1)$$

$$P_{in} = G \times A_{pv} \qquad (2)$$

where, $I_{pv,max}$ is the maximum operating current, $V_{pv,max}$ is the maximum operating voltage, $P_{in}$ is the energy value of the incident rays, $G$ is the solar irradiance and $A_{pv}$ is the area of the PV panel.

**Thermoelectric generator.** For the TEG, the efficiency can be obtained using Eq. (3)

$$\eta_{TEG} = \frac{P_{TEG}}{Q} = \frac{I_{TEG,max} \times V_{TEG,max}}{Q} \qquad (3)$$

where, $I_{TEG,max}$ is the maximum operating current, $V_{TEG,max}$ is the maximum operating voltage and $Q$ is the net heat flow into the TEG from the backside of the PV panel.

Q is mathematically represented in Eq. ( 4 ).

$$Q = \frac{K A \Delta T}{d} \qquad (4)$$

where $K$ is the thermal conductivity, $A$ is the area of the TEG, $\Delta T$ is the temperature difference between the TEG hot and cold sides, and $d$ is the thickness of the TEG.

The combined efficiency of the system, $\eta_{sys}$ can then be computed using Eq. ( 5 ).

$$\eta_{sys} = \frac{P_{pv} + P_{TEG}}{P} \qquad (5)$$

**Results and Discussion**

In this section, we present and discuss the acquired data as well as the experimental outcomes. We also analyze the performance variations observed across the different configurations employed in the study.

Solar irradiance increased steadily from 990 W/m² at 10:30 AM to a peak of 1084 W/m² at 11:30 AM, then gradually declined to 200 W/m² by 6:00 PM, as shown in Fig. 3.

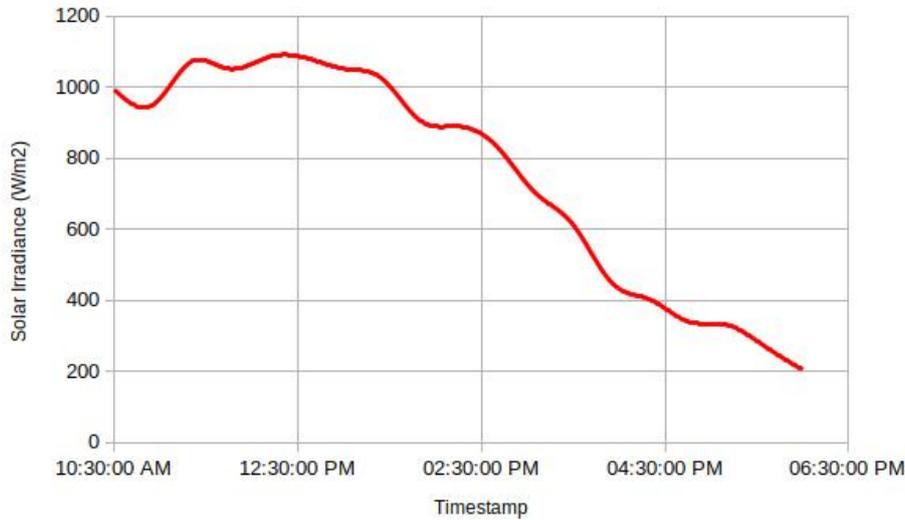

Figure 3: Solar Irradiance against time from 10:30:00 AM to 06:00:00 PM over a 25-minute interval

The PV power outputs of four configurations under varying timestamps (Fig. 4a) and solar irradiance (Fig. 4b) showed similar daily trends, with outputs peaking around midday and decreasing towards evening, consistent with typical solar behavior. Among the configurations, PV-PCM produced the highest power, peaking near midday, followed closely by PV-PCM-TEG. The difference in power output between configurations was most pronounced at peak irradiance and narrowed towards sunset, highlighting the role of PCMs in maximizing output during high solar intensity. PCMs absorb heat during peak irradiation, lowering the panel temperature and enhancing efficiency by reducing temperature-induced losses. They also release stored heat during cooler periods, mitigating power fluctuations. In contrast, the standalone PV system had the lowest and most variable power output across all timestamps, demonstrating greater susceptibility to temperature fluctuations. The PV-TEG-PCM configuration showed higher power output than the standalone system but slightly less than PV-PCM and PV-PCM-TEG. While TEGs effectively removed excess heat, maintaining lower PV cell temperatures and enhancing output, they lack the buffering capability of PCMs to absorb and release heat. This explains why PV-PCM outperformed PV-TEG, stabilizing cell temperatures and ensuring more consistent power generation.

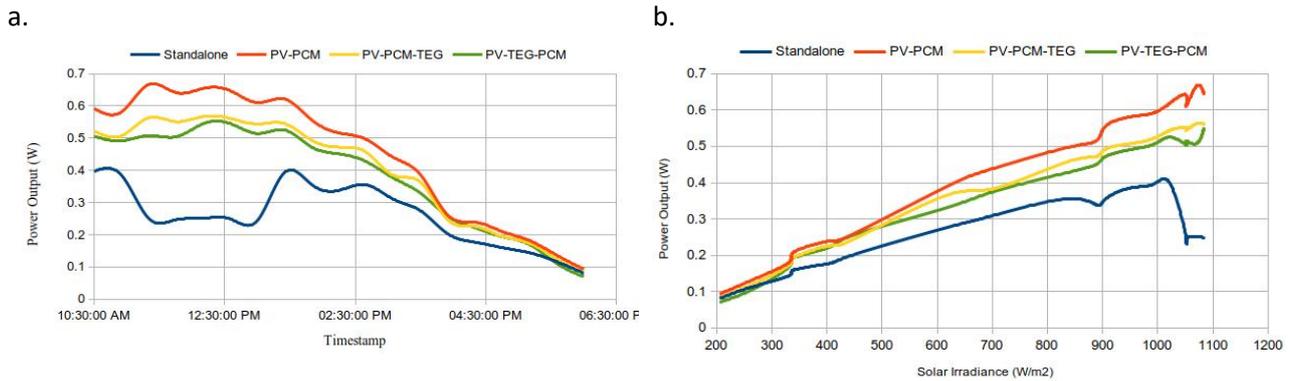

Figure 4: Solar PV power output at (a) various times of the day; (b) varying solar irradiance levels

Fig. 5a and Fig. 5b show PV efficiency trends for different configurations across varying timestamps and solar irradiance, respectively. The standalone system exhibited greater fluctuations, with efficiency dropping notably beyond 1 kW/m² due to increased cell temperatures from excess heat. This aligns with the Shockley-Queisser limit, where efficiency peaks at moderate irradiances but declines at higher levels due to thermal losses [30]. High irradiance levels observed midday (11:00 AM – 1:00 PM) coincided with efficiency drops in the standalone PV system. In contrast, PV-PCM and PV-PCM-TEG maintained higher efficiency throughout, particularly during peak sun hours, as PCMs absorbed excess heat, preventing extreme panel temperatures and sustaining efficiency.

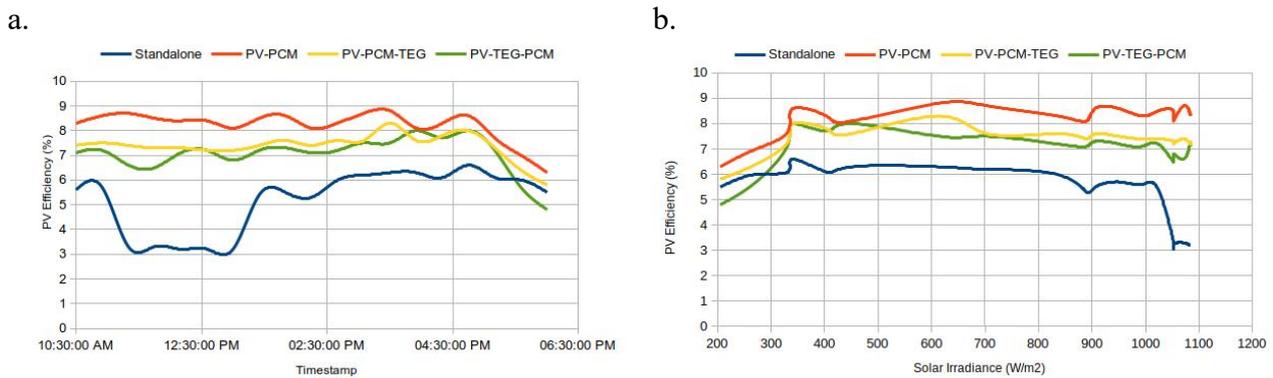

Figure 5: Solar PV cell efficiency at (a) various times of the day; (b) varying solar irradiance levels

Fig. 6 shows the average PV efficiency of the four configurations, all of which outperform the standalone system. The PV-PCM configuration achieved a 33.33% higher efficiency, PV-PCM-TEG had a 25.76% increase, and PV-TEG-PCM showed a 21.21% improvement compared to the standalone system.

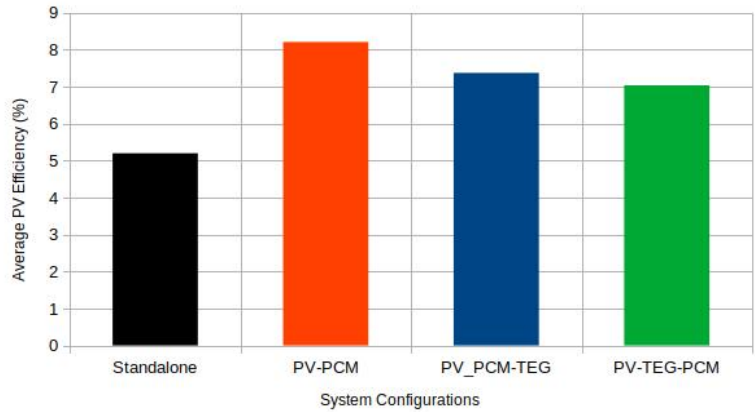

Figure 6: Average PV Efficiency (in percent) for four different system configurations.

Fig. 7 and Fig. 8 show the TEG power output for PV-PCM-TEG and PV-TEG-PCM configurations across various timestamps and solar irradiance levels respectively. Both configurations demonstrate a positive correlation between solar irradiance and TEG power, with PV-PCM-TEG outperforming PV-TEG-PCM across all conditions. The PV-PCM-TEG configuration shows a power range of 0.0001W to 0.003W, maintaining higher output even at lower irradiances, while PV-TEG-PCM ranges from 0W to 0.0013W and drops to zero below 403 W/m². The non-linear power increase at higher irradiances suggests limitations due to TEG internal resistances and thermal saturation. The placement of PCM between the PV cell and TEG enhances heat transfer, reduces thermal losses, and provides a stable temperature gradient, resulting in consistently higher TEG power output for the PV-PCM-TEG configuration.

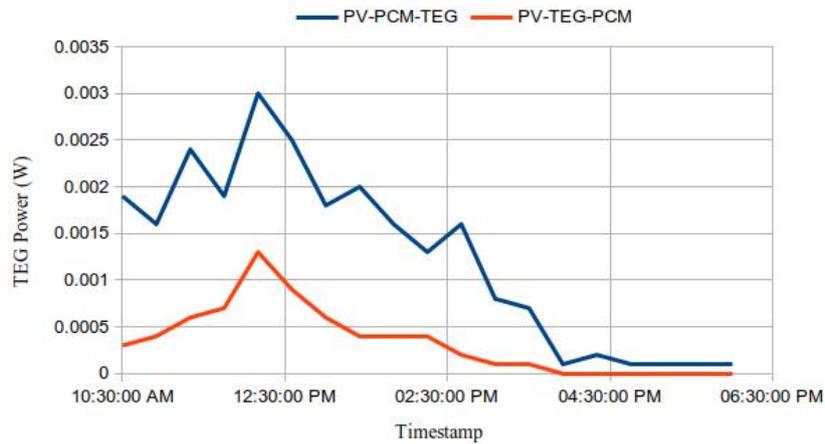

Figure 7. TEG power output at various times of the day

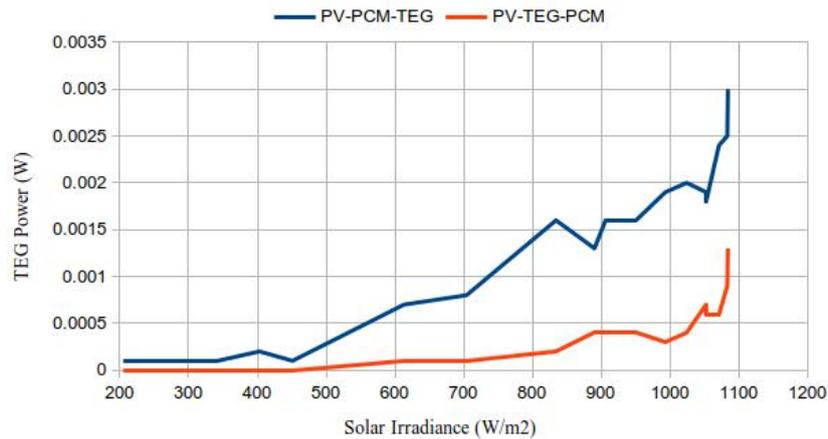

Figure 8. TEG power output at varying solar irradiance levels

**Conclusion and Future Studies**

This study investigated the integration of thermoelectric generators (TEGs) and phase change materials (PCMs) to enhance solar PV efficiency. Experimental results under real-world conditions demonstrated that hybrid configurations—PV-PCM, PV-PCM-TEG, and PV-TEG-PCM—outperformed the standalone PV system. Key findings include:

1. The PV-PCM system showed a 68.04% increase in power output, while PV-PCM-TEG and PV-TEG-PCM achieved 43.06% and 37.51% improvements, respectively, highlighting the potential of integrated systems to significantly enhance power generation.
2. Electrical efficiency gains were also significant, with increases of 33.33% for PV-PCM, 25.76% for PV-PCM-TEG, and 21.21% for PV-TEG-PCM compared to the standalone PV system, underscoring the value of TEG and PCM integration in improving energy conversion efficiency.

The study confirmed that incorporating PCM at the back of the PV panel effectively reduces panel temperature, enhancing overall power output and efficiency. Among the tested configurations, the PV-PCM setup delivered the best performance, establishing the advantage of hybrid approaches in managing heat and enhancing PV efficiency. These results underscore the potential of TEG and PCM technologies to address heat accumulation in solar panels, providing valuable insights for optimizing solar energy systems.

Moving forward, we recommend further exploration of diverse TEG designs, materials, and PCM types for large-scale applications. Future research should also investigate the impact of variables such as humidity and solar intensity fluctuations, assess the long-term durability and economic viability of TEG-PCM systems, and explore integration with energy storage to fully realize their potential in sustainable energy solutions.